\documentclass{article}
\usepackage{spconf,amsmath,graphicx}
\usepackage{geometry}
\usepackage{color}
\usepackage{cite}
\usepackage{amsmath}
\usepackage{array}
\usepackage{fancyhdr}
\usepackage{latexsym}
\usepackage{epsfig}
\usepackage{amssymb}
\usepackage{amsfonts}
\usepackage{xspace}
\usepackage{makeidx}
\usepackage{graphics}
\usepackage{theorem}
\usepackage{subfigure}
\usepackage{multirow}
\usepackage{longtable}
\usepackage{tabularx}
\usepackage{algorithm}
\usepackage{algorithmicx}
\usepackage{algpseudocode}
\DeclareMathOperator*{\minimize}{minimize}

\newtheorem{theorem}{\textbf{Theorem}}

\title{Hybrid Beamforming for Single Carrier MmWave MIMO Systems}
\name{Tian Lin \qquad Jiaqi Cong \qquad  Yu Zhu\thanks{\noindent This work was supported by National Natural Science Foundation of China under Grant No. 61771147 and No. 61271223. (Corresponding author: Yu Zhu.)}}
\address{School of Information Science and Technology\\Fudan University, Shanghai, China\\
Email: lint17, jqcong16, zhuyu@fudan.edu.cn}
\begin{document}
%
\maketitle
\begin{abstract}
Hybrid analog and digital beamforming (HBF) has been recognized as an attractive technique offering a tradeoff between hardware implementation limitation and system performance for future broadband millimeter wave (mmWave) communications. In contrast to most current works focusing on the HBF design for orthogonal frequency division multiplexing based mmWave systems, this paper investigates the HBF design for single carrier (SC) systems due to the advantage of low peak-to-average power ratio in transmissions. By applying the alternating minimization method, we propose an efficient HBF scheme based on the minimum mean square error criterion.   Simulation results show that the proposed scheme outperforms the conventional HBF scheme for SC systems.
\end{abstract}
\section{Introduction}\label{sec:intro}
Hybrid analog and digital beamforming (HBF) architecture has emerged as an attractive technique to significantly reduce the hardware cost and power consumption while providing sufficient beamforming gains for millimeter wave (mmWave) systems with large number of antennas \cite{Heath2014,Alkhateeb2014}. Given the large bandwidth, HBF should be combined with broadband transmission technologies such as orthogonal frequency division multiplexing (OFDM) \cite{YUWEI2017,Zhangjun2016}. However, the intrinsic problem of OFDM such as high peak-to-average power ratio (PAPR)  is inherited in such systems and may become more serious due to the high implementation cost of mmWave devices.

Nevertheless, single carrier (SC) with frequency domain equalization (FDE) has been regarded as an alternative technology to OFDM due to its  advantage of lower PAPR \cite{Buzzi2018,Xiao2015,Li2017}. The analog beamforming for SC mmWave systems with only one radio frequency (RF) chain has been investigated in \cite{Xiao2015,Li2017} with different design criteria. More recently, the authors have studied the HBF design for SC mmWave systems with multiple RF chains \cite{Buzzi2018}. However, the HBF was optimized with the consideration of only one channel path with the largest gain.

In this paper, we investigate the HBF design for SC broadband mmWave systems utilizing all channel information and aiming at minimizing the sum of the mean square errors (sum-MSE) of the multiple data streams. 
To solve this non-convex HBF problem, we first consider the case where the number of RF chains is equal to that of data streams, and derive the optimal digital combiners at each frequency tone. The digital precoder is optimized with the help of an auxiliary orthogonal constraint to simplify the problem. With these optimized digital beamformers, the analog ones can be optimized iteratively based on the alternating minimization method \cite{Zhangjun2016}.  We further generalize the HBF design to the case of more RF chains. Simulation results show that the proposed HBF scheme achieves significant performance improvement over the  conventional beamforming schemes for SC mmWave systems.

\begin{figure*}[!t]
\centering
\includegraphics[width=4.8in]{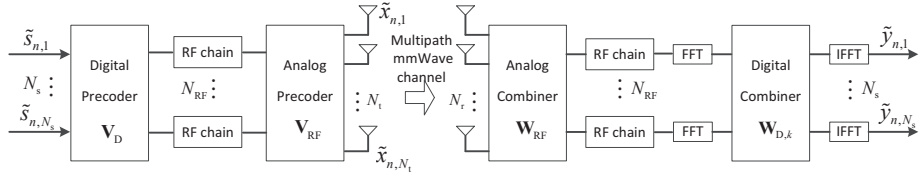}
\caption{Diagram of a point-to-point SC broadband mmWave MIMO system with HBF.}
\label{fig:sys-narrowband}
\end{figure*}

\section{System Model}\label{sec:format}
Consider a point-to-point broadband SC mmWave system in Fig. 1, where the transmitter (Tx) with $N_\mathrm{t}$ antennas and $N_\mathrm{RF}$ RF chains sends $N_\mathrm{s}$ data streams to the receiver (Rx) with $N_\mathrm{r}$ antennas and $N_{\mathrm{RF}}$ RF chains. To address the hardware limitation challenge, it is assumed that  $\min (N_\mathrm{t},N_\mathrm{r}) \gg  N_\mathrm{RF} = N_\mathrm{s}$\footnote{The extension to the general case of $N_\mathrm{RF} > N_\mathrm{s}$ will be addressed in Section \ref{sec:general}.}. Assume that each block of a data stream contains $N$ independent quadrature amplitude modulation (QAM) symbols. Denote the symbols of all the data streams at time $n$ by an $N_s\times 1$ vector  $\widetilde{\mathbf{s}}_{n}$, for $n=0,1,\ldots,N-1$, with $E\{\widetilde{\mathbf{s}}_{n}\widetilde{\mathbf{s}}_{n}^H\} = \mathbf{I}_{N_\mathrm{s}}$, where $E\{\cdot\}$ and $\{\cdot\}^H$ denote the expectation and the conjugate transpose, respectively, and $\mathbf{I}_{N_\mathrm{s}}$ denotes the $N_\mathrm{s}\times N_\mathrm{s}$ identity matrix. After inserting a cyclic prefix (CP) for each block to deal with the channel frequency selective fading \cite{Li2017},  the Tx multiples $\widetilde{\mathbf{s}}_{n}$ by an ${N_\mathrm{RF}}\times{N_\mathrm{s}}$ digital precoding matrix  $\mathbf{V}_\mathrm{D}$ and then up-converts the precoded signals to the carrier frequency through $N_\mathrm{RF}$ RF chains. To guarantee a low PAPR in SC transmissions,  it is assumed that $\mathbf{V}_\mathrm{D}$ is invariant for the whole available bandwidth \cite{Buzzi2018}. Finally, an ${N_\mathrm{t}}\times{N_\mathrm{RF}}$ analog precoding matrix $\mathbf{V}_\mathrm{RF}$ is implemented  using phase shifters to construct the signals for the transmit antenna array. From the equivalent baseband point of view, the transmitted signal can be represented by $\widetilde{\mathbf{x}}_{n} = \mathbf{V}_\mathrm{RF} \mathbf{V}_\mathrm{D} \widetilde{\mathbf{s}}_{n}$.

At the Rx side, the HBF is processed in the opposite way of that at the Tx side. That is, first by an $N_\mathrm{r}\times N_\mathrm{RF}$ analog combining matrix $\mathbf{W}_\mathrm{RF}$ and then by a number of digital matrices with the size $N_\mathrm{RF} \times N_\mathrm{s}$ for different frequency tones. Note that, unlike that in the design of the digital precoder, which is same for the whole band to guarantee a low PAPR, the digital combining matrices can be optimized for different frequency tones. With the help of CP, the signal vector after receive HBF at the $k$th frequency tone, for $k=0,\ldots,N-1$, can be represented by
\begin{equation}\label{eqn:yk}
{\mathbf{y}}_k={\mathbf{W}^H_{\mathrm{D},k}}{\mathbf{W}^H_\mathrm{RF}}{\mathbf{H}}_{k}{\mathbf{V}_\mathrm{RF}} {\mathbf{V}_\mathrm{D}}{\mathbf{s}}_k+{\mathbf{W}^H_{\mathrm{D},k}}{\mathbf{W}^H_\mathrm{RF}}
{\mathbf{u}}_k, 
\end{equation}
where $\mathbf{s}_k = \frac{1}{\sqrt{N}}\sum_{n=0}^{N-1}\widetilde{\mathbf{s}}_n e^{-j\frac{2\pi}{N}nk}$, $\mathbf{H}_k$ and $\mathbf{u}_k$ denote the frequency component of the transmitted signal, the channel frequency response, and the additive noise vector at the $k$th frequency tone, respectively.  $\mathbf{W}_{\mathrm{D},k}$ denotes the digital combiner matrix at the $k$th frequency tone. It is assumed that $\mathbf{u}_k$ satisfies the circularly symmetric complex Gaussian distribution with zero mean and covariance matrix ${\sigma^2}\mathbf{I}_{N_r}$. Define the processed signal vector in the time domain after inverse Fourier transform (IFFT) as $\widetilde{\mathbf{y}}_n = \frac{1}{{\sqrt N }}\sum_{k = 0}^{N - 1} \mathbf{y}_k{e^{j\frac{2\pi}{N} nk}}$. In this study, we take the streams' sum-MSE as the HBF optimization objective, which is given by
\begin{equation}\label{MSE1}
\begin{split}
J &=  \mathrm{tr}\left(E\{(\widetilde{\mathbf{y}}_n -\widetilde{\mathbf{s}}_n)(\widetilde{\mathbf{y}}_n-\widetilde{\mathbf{s}}_n)^H\} \right)\\
&= \frac{1}{N}\sum\limits_{k = 0}^{N - 1} (\|\mathbf{I}_{N_\mathrm{s}}-\mathbf{G}_k\|_F^2+\sigma^2\|{\mathbf{W}_\mathrm{RF}}{\mathbf{W}_{\mathrm{D},k}}\|_F^2),
\end{split}
\end{equation}
where $\mathbf{G}_k\triangleq {\mathbf{W}^H_{\mathrm{D},k}}{\mathbf{W}^H_\mathrm{RF}}{\mathbf{H}}_{k}{\mathbf{V}_\mathrm{RF}}{\mathbf{V}_\mathrm{D}}$. $\mathrm{tr}(\cdot)$ and $\left\| \cdot \right\|_F$ respectively denote the trace and the Frobenius norm of a matrix. Then, the HBF optimization problem is formulated as follows
\begin{equation}\label{pro:opt}
\begin{array}{cl}
\displaystyle{\minimize_{{\mathbf{V}_\mathrm{RF}},{\mathbf{V}_\mathrm{D}},{\mathbf{W}_\mathrm{RF}},{\mathbf{W}_{\mathrm{D},k}}}} & J \\
\mathrm{subject \; to} & \|{\mathbf{V}_\mathrm{RF}}{\mathbf{V}_\mathrm{D}}\|_F^2\leq 1; \\ &{\mathbf{V}_\mathrm{RF}}\in \mathcal{V}; \;\;\;\;{\mathbf{W}_\mathrm{RF}}\in \mathcal{W},
\end{array}
\end{equation}
where $\|{\mathbf{V}_\mathrm{RF}}{\mathbf{V}_\mathrm{D}}\|_F^2\leq 1$ represents  the maximum  transmit power (normalized to 1) constraint,  and $\mathcal{V}$ and $\mathcal{W}$ denote the sets of feasible analog precoders and combiners induced by the constant modulus constraint of the phase shifters.

\section{HBF Design}\label{sec:pagestyle}
In this section, we first optimize the low-dimension digital beamformers $\mathbf{W}_{\mathrm{D},k}$ and $\mathbf{V}_{\mathrm{D}}$, and then deal with the intractable constant modulus constraint for the solution of analog beamformers $\mathbf{W}_{\mathrm{RF}}$ and $\mathbf{V}_{\mathrm{RF}}$. We finally extend the HBF optimization to the general case of $N_\mathrm{RF} > N_\mathrm{s}$.

\subsection{Digital Beamformers Design}
By fixing $\mathbf{V}_{\mathrm{D}}$, $\mathbf{V}_{\mathrm{RF}}$ and $\mathbf{W}_{\mathrm{RF}}$ in (\ref{MSE1}), the optimal $\mathbf{W}_{\mathrm{D},k}$ can be obtained by differentiating $J$ in (\ref{MSE1}) with $\mathbf{W}_{\mathrm{D},k}$ and setting the result to zero\footnote{ ${\mathbf{W}_{\mathrm{D},k}}$ can also be regarded as the coefficient at the $k$th frequency tone of the linear frequency domain equalizer based on the MMSE criterion.}. That is,
\begin{equation}\label{eqn:Wk}
{\mathbf{W}_{\mathrm{D},k}}=( {\mathbf{B}_{k}}{\mathbf{B}}_{k}^H+\sigma^2\mathbf{A}^{-1})^{-1}{\mathbf{B}}_{k}.
\end{equation}
where $\mathbf{A}\triangleq ({\mathbf{W}^H_\mathrm{RF}}{\mathbf{W}_\mathrm{RF}})^{-1}$ and ${\mathbf{B}}_{k}\triangleq {\mathbf{W}^H_\mathrm{RF}}{\mathbf{H}}_{k} {\mathbf{V}_\mathrm{RF}}{\mathbf{V}_\mathrm{D}}$. By substituting (\ref{eqn:Wk}) into (\ref{MSE1}), we have
\begin{equation}\label{eqn:MSE1}	
   J=\frac{1}{N}\mathrm{tr}\sum\limits_{k = 0}^{N - 1} \left(\frac{1}{{\sigma ^2}}{\mathbf{B}}_{k}^H \mathbf{A} {\mathbf{B}_{k}}+ {\mathbf{I}}_{N_s}\right)^{-1}.
\end{equation}
Note that the optimal $\mathbf{W}_{\mathrm{D},k}$ is now a function of $\mathbf{V}_\mathrm{D}$, $\mathbf{V}_\mathrm{RF}$ and $\mathbf{W}_\mathrm{RF}$ which can be further optimized based on (\ref{eqn:MSE1}). However, it can be seen from (\ref{eqn:MSE1}) that it is very hard to derive the optimal $\mathbf{V}_\mathrm{D}$, $\mathbf{V}_\mathrm{RF}$, $\mathbf{W}_\mathrm{RF}$ and some algorithms are needed to find a suboptimal solution.

To optimize the digital precoder $\mathbf{V}_\mathrm{D}$, it is worth noting that the dimension (which determines the optimization freedom) of $\mathbf{V}_\mathrm{D}$ is much lower than that of  $\mathbf{V}_\mathrm{RF}$ or $\mathbf{W}_\mathrm{RF}$. Thus, from the complexity point of view, a simplification is made for the optimization of $\mathbf{V}_\mathrm{D}$. Inspired by the method in \cite{Zhangjun2016}, an  auxiliary orthogonal constraint is added to $\mathbf{V}_\mathrm{D}$. That is,
\begin{equation}\label{VDVDH}
\mathbf{V}^H_\mathrm{D}\mathbf{V}_\mathrm{D} = \gamma\mathbf{V}^H_\mathrm{U}\mathbf{V}_\mathrm{U}=\gamma \mathbf{I}_{N_\mathrm{s}},
\end{equation}
where $\gamma>0$ and $\mathbf{V}_\mathrm{U}$ is a para-unitary matrix with $\mathbf{V}_\mathrm{U}^H\mathbf{V}_\mathrm{U} = \mathbf{I}_{N_\mathrm{s}}$. This constraint  implies that  the columns of $\mathbf{V}_\mathrm{D}$ should be mutually orthogonal.  From the assumption of  $N_\mathrm{s}=N_\mathrm{RF}$ made in Section \ref{sec:format}, it can be found that $\mathbf{V}_\mathrm{D}\mathbf{V}^H_\mathrm{D} =\gamma \mathbf{I}_{N_\mathrm{s}}$, and the eigenvalues of $\left(\frac{1}{{\sigma ^2}}{\mathbf{B}}_{k}^H \mathbf{A} {\mathbf{B}_{k}}+ {\mathbf{I}}_{N_s}\right)$ are the same as those of  $\left(\frac{\gamma}{{\sigma ^2}}{\mathbf{C}}_{k}^H \mathbf{A} {\mathbf{C}_{k}}+ {\mathbf{I}}_{N_\mathrm{RF}}\right)$, where
\begin{equation}\label{Ck}
  {\mathbf{C}}_{k}\triangleq {\mathbf{W}^H_\mathrm{RF}}{\mathbf{H}}_{k}{\mathbf{V}_\mathrm{RF}}.
  \end{equation}
Thus, the objective $J$ in (\ref{eqn:MSE1}) can be represented by
\begin{equation}\label{MSE3}
  J=\frac{1}{N}\mathrm{tr}\sum_{k = 0}^{N - 1} \left(\frac{\gamma}{{\sigma ^2}}{\mathbf{C}}_{k}^H \mathbf{A} {\mathbf{C}_{k}}+ {\mathbf{I}}_{N_\mathrm{RF}}\right)^{-1},
\end{equation}
which is not associated with  $\mathbf{V}_\mathrm{U}$ as long as $\mathbf{V}_\mathrm{U}$ is a unitary matrix. This greatly simplifies the complexity in optimizing $\mathbf{V}_\mathrm{D}$.  To further optimize $\gamma$, it can be seen from (\ref{MSE3}) that $\gamma$ should be set as large as possible. However, as limited by the maximum transmit power constraint in (\ref{pro:opt}),   the largest $\gamma$ is  $\frac{1}{N_\mathrm{t}N_\mathrm{s}}$, which can be used to form the final $\mathbf{V}_\mathrm{D}$.

\subsection{Analog Beamformers Design}\label{PrecodingEVD}
Directly minimizing the objective of (\ref{MSE3}) to obtain the analog beamformers is still difficult.  According to \cite{YUWEI2016}, it can be approximated that $\mathbf{W}^H_\mathrm{RF}\mathbf{W}_\mathrm{RF}\approx N_\mathrm{r}\mathbf{I}_{N_\mathrm{RF}}$ and $\mathbf{V}^H_\mathrm{RF}\mathbf{V}_\mathrm{RF}\approx N_\mathrm{t}\mathbf{I}_{N_\mathrm{RF}}$ for large-scale MIMO systems,  as the optimized analog beamforming vectors for different streams are likely orthogonal to each other.  Therefore, we add two reasonable constraints of $\mathbf{W}^H_\mathrm{RF}\mathbf{W}_\mathrm{RF}= N_\mathrm{r}\mathbf{I}_{N_\mathrm{RF}}$ and $\mathbf{V}^H_\mathrm{RF}\mathbf{V}_\mathrm{RF}= N_\mathrm{t}\mathbf{I}_{N_\mathrm{RF}}$. Aiming at minimizing (\ref{MSE3}) and omitting the scalar $\frac{1}{N}$, we can simplify the original optimization for the analog beamformers as follows
\begin{equation}\label{pro:opt2}
\begin{array}{cl}
\displaystyle{\minimize_{{\mathbf{V}_\mathrm{RF}},{\mathbf{W}_\mathrm{RF}}}} & \sum\limits_{k = 0}^{N - 1}\mathrm{tr} \left(\frac{\gamma}{N_\mathrm{r}\sigma^2}{\mathbf{C}}^H_{k}{\mathbf{C}_{k}} + {\mathbf{I}}_{N_\mathrm{RF}}\right)^{-1}\\
\mathrm{subject \; to} & {\mathbf{V}_\mathrm{RF}}\in \mathcal{V} \quad {\mathbf{W}_\mathrm{RF}}\in \mathcal{W} \\
& \mathbf{W}^H_\mathrm{RF}\mathbf{W}_\mathrm{RF}= N_\mathrm{r}\mathbf{I}_{N_\mathrm{RF}},\;\mathbf{V}^H_\mathrm{RF}\mathbf{V}_\mathrm{RF}= N_\mathrm{t}\mathbf{I}_{N_\mathrm{RF}}.
\end{array}\nonumber
\end{equation}
Based on the method  of alternating minimization \cite{Zhangjun2016}, we decompose this problem into two subproblems where the one is to optimize $\mathbf{V}_\mathrm{RF}$ with a fixed $\mathbf{W}_\mathrm{RF}$ and the other is to optimize $\mathbf{W}_\mathrm{RF}$ with an updated $\mathbf{V}_\mathrm{RF}$. It is worth noting that since $\mathrm{tr}(\mathbf{Z}^H\mathbf{Z}+\mathbf{I})^{-1}= \mathrm{tr}(\mathbf{Z}\mathbf{Z}^H+\mathbf{I})^{-1}$ for a square matrix $\mathbf{Z}$, it is not difficult to find that the two subproblems can be unified into the same form and thus optimized exactly with the same method. In the following, we take the subproblem of optimizing $\mathbf{V}_\mathrm{RF}$ with a fixed $\mathbf{W}_\mathrm{RF}$ as an example.

We first temporarily ignore the constant modulus constraint for simplification. However, the simplified problem is still difficult to solve. Nevertheless, instead of dealing with the objective directly, we derive its upper bound with the help of the following theorem.
\begin{theorem}\label{Theorem:1}
Consider an  $m\times m$ positive definite and Hermitian matrix $\mathbf{M}$ and an arbitrary $m\times n$  $(m>n)$  para-unitary matrix  $\mathbf{R}$, i.e., $\mathbf{R}^H\mathbf{R}=\mathbf{I}_n$. Let $\mu_1,...,\mu_{n}$ and $\lambda_1,...,\lambda_{n}$ denote the eigenvalues in descending order of $(\mathbf{R}^H\mathbf{M}\mathbf{R})^{-1}$ and $\mathbf{R}^H\mathbf{M}^{-1}\mathbf{R}$, respectively,  it can be proved that $\mu_k \le \lambda_k, \forall k$.
\end{theorem}
\textit{Proof}: According to Courant-Fisher min-max theorem \cite{1990Matrix},
\begin{equation}
\lambda_k = \mathop {\max }\limits_\mathbb{U} \mathop {\min }\limits_{\mathbf{x} \in \mathbb{U}}
\frac{\mathbf{x}^H\mathbf{R}^H\mathbf{M}^{-1}\mathbf{R}\mathbf{x}}{\mathbf{x}^H\mathbf{x}}
=\mathop {\max }\limits_\mathbb{U} \mathop {\min }\limits_{\mathbf{x} \in \mathbb{F}}
\frac{\mathbf{x}^H\mathbf{M}^{-1}\mathbf{x}}{\mathbf{x}^H\mathbf{x}}, \nonumber
\end{equation}
where $\mathbf{x}$ is a non-zero vector, $\mathbb{U}$ denotes a $k$-dimension subspace of $\mathbb{C}^m$ and $\mathbb{F}$ is a new subspace after a linear transform of $\mathbf{R}$ to $\mathbb{U}$. Similarly, as $1{\rm{/}}\mu_k$ can be regarded as the $(N-k+1)$th largest eigenvalue of $\mathbf{R}^H\mathbf{M}\mathbf{R}$, we  have
\begin{equation}
\frac{1}{\mu_k} = \mathop {\min }\limits_\mathbb{U} \mathop {\max }\limits_{\mathbf{x} \in \mathbb{U}}
\frac{\mathbf{x}^H\mathbf{R}^H\mathbf{M}\mathbf{R}\mathbf{x}}{\mathbf{x}^H\mathbf{x}}\mathop {\min }\limits_\mathbb{U} \mathop {\max }\limits_{\mathbf{x} \in \mathbb{F}}
\frac{\mathbf{x}^H\mathbf{M}\mathbf{x}}{\mathbf{x}^H\mathbf{x}}.
\end{equation}
Then,
\begin{equation}
\mu_k = \mathop {\max }\limits_\mathbb{U} \mathop {\min }\limits_{\mathbf{x} \in \mathbb{F}}
\frac{\mathbf{x}^H\mathbf{x}}{\mathbf{x}^H\mathbf{M}\mathbf{x}}.
\end{equation}
According to the fact that $\mathbf{M}$ is positive definite and by using the Jensen's inequality, $\frac{\mathbf{x}^H\mathbf{x}}{\mathbf{x}^H\mathbf{M}\mathbf{x}}\le \frac{\mathbf{x}^H\mathbf{M}^{-1}\mathbf{x}}{\mathbf{x}^H\mathbf{x}}$ holds for any non-zero vector $\mathbf{x}$. Thus, the proof is completed. \hfill $\square$

By setting $\mathbf{M}_k \triangleq \left(\frac{\gamma}{N_\mathrm{r}\sigma^2}\mathbf{H}^H_{k}\mathbf{W}_\mathrm{RF}\mathbf{W}^H_\mathrm{RF}\mathbf{H}_{k}  + {\mathbf{I}}_{N_t}\right)$ and using Theorem 1, we can derive that
\begin{equation}\label{derive4}
\begin{split}
\sum_{k = 0}^{N - 1}\mathrm{tr} \left(\frac{\gamma}{N_\mathrm{r}\sigma^2} {\mathbf{C}}^H_{k}{\mathbf{C}_{k}} + {\mathbf{I}}_{N_\mathrm{RF}}\right)^{-1}=\sum\limits_{k = 0}^{N - 1}\mathrm{tr} \left(\mathbf{V}_\mathrm{RF}^H\mathbf{M}_\mathrm{k}\mathbf{V}_\mathrm{RF}\right)^{-1}\\  \le \sum\limits_{k = 0}^{N - 1}\mathrm{tr} \left(\mathbf{V}_\mathrm{RF}^H \mathbf{M}_\mathrm{k}^{-1}\mathbf{V}_\mathrm{RF}\right)
 =\mathrm{tr}\left(\mathbf{V}_\mathrm{RF}^H \left(\sum\limits_{k = 0}^{N - 1}\mathbf{M}_\mathrm{k}^{-1}\right)\mathbf{V}_\mathrm{RF}\right). \nonumber
\end{split}
\end{equation}
According to \cite{KFAN}, it can be proved that the optimal $\mathbf{V}_\mathrm{RF}$ to minimize the above upper bound is $\sqrt{N_\mathrm{t}}$ times the isometric matrix containing the $N_\mathrm{RF}$ eigenvectors associated with the smallest $N_\mathrm{RF}$  eigenvalues of $\sum_{k = 0}^{N - 1} \mathbf{M}_{k}^{-1}$, which can be obtained through eigenvalue decomposition (EVD). To further take the constant modulus constraint into account, a simple but effective way to form the optimized $\mathbf{V}_\mathrm{RF}$ is to directly extract the phase of each elements of these eigenvectors.

The above EVD-based scheme  can be applied to solve the subproblem for optimizing $\mathbf{W}_\mathrm{RF}$. Finally, using the alternating minimization method  and by performing iterations between the optimization of $\mathbf{V}_\mathrm{RF}$ and that of $\mathbf{W}_\mathrm{RF}$ until a stop condition is satisfied, a pair of optimized analog beamformers are obtained.
\begin{algorithm}[t]
\caption{The EVD-HBF Scheme}
\begin{algorithmic}[1]
\State Initialize $\mathbf{V}_{\mathrm{RF},\,0} $ and $\mathbf{W}_{\mathrm{RF},\,0} $ randomly and set $i=0$;
\State \textbf{repeat}	
\State $i\leftarrow i+1$;
\State Update $\mathbf{V}_{\mathrm{RF},\, i}$ using the fixed $\mathbf{W}_{\mathrm{RF},\, i-1}$, and then update $\mathbf{W}_{\mathrm{RF},\, i}$ using the fixed $\mathbf{V}_{\mathrm{RF},\, i}$ via the  EVD based algorithm in Section \ref{PrecodingEVD};
\State \textbf{Until} a stopping condition is satisfied;
\State $\mathbf{V}_{\mathrm{RF}}=\mathbf{V}_{\mathrm{RF},\, i}$, $\mathbf{W}_{\mathrm{RF}}=\mathbf{W}_{\mathrm{RF},\, i}\,$;
\State Obtain $\mathbf{V}_\mathrm{U}$ with fixed analog beamformers using the EVD based algorithm;
\State $\mathbf{V}_\mathrm{D}=(\mathrm{tr}({\mathbf{V}_\mathrm{RF}}{\mathbf{V}_\mathrm{U}} {\mathbf{V}^H_\mathrm{U}}{\mathbf{V}^H_\mathrm{RF}}))^{-\frac{1}{2}}\times \mathbf{V}_\mathrm{U}$;
\State Compute $\mathbf{W}_\mathrm{D}$   according to (\ref{eqn:Wk}).
\end{algorithmic}\label{alg:manifold}
\end{algorithm}

\subsection{HBF Design for General Cases}\label{sec:general}
For the more general case of $N_\mathrm{RF}>N_\mathrm{s}$, the optimization of $\mathbf{W}_{\mathrm{D},k}$ is the same as that in (\ref{eqn:Wk}) and the difference is in that of other beamformers. First, still add an orthogonal constraint to $\mathbf{V}_\mathrm{D}$. That is, $\mathbf{V}^H_\mathrm{D}\mathbf{V}_\mathrm{D} = \gamma \mathbf{I}_{N_\mathrm{s}}$, it can be proved from the Weyl Theorem \cite{1990Matrix} that the  descending order eigenvalues of $\gamma{\mathbf{C}}^H_{k} \mathbf{A}{\mathbf{C}_{k}}$  is item by item larger than that of ${\mathbf{B}}^H_{k} \mathbf{A}{\mathbf{B}_{k}}$, where the definitions of $\mathbf{A}$, ${\mathbf{B}_{k}}$, and ${\mathbf{C}_{k}}$ can be found from (\ref{eqn:Wk}) and (\ref{Ck}), respectively. Then, using the relationship between a matrix' trace and eigenvalues,  we can derive a lower bound of (\ref{eqn:MSE1}) as follows
\begin{equation}\label{JLB}
J_\mathrm{L}=\frac{1}{N}\sum_{k = 0}^{N - 1}\left(\mathrm{tr} \left(\frac{\gamma}{\sigma^2}{\mathbf{C}}^H_{k} \mathbf{A}{\mathbf{C}_{k}} + {\mathbf{I}}_{N_\mathrm{RF}}\right)^{-1}+N_\mathrm{s}-N_\mathrm{RF}\right).
\end{equation}
As $J_\mathrm{L}$ has a similar form to the objective function of (\ref{MSE3}) after omitting the constant terms, the proposed algorithm in Section \ref{PrecodingEVD} can be applied to form  $\mathbf{V}_\mathrm{RF}$ and $\mathbf{W}_\mathrm{RF}$.
Finally, to optimize $\mathbf{V}_\mathrm{D}=\gamma \mathbf{V}_\mathrm{U}$ with   the optimized $\mathbf{V}_\mathrm{RF}$ and $\mathbf{W}_\mathrm{RF}$, by substituting the orthogonal constraint (\ref{VDVDH}) and the assumption $\mathbf{A}\triangleq(\mathbf{W}^H_\mathrm{RF}\mathbf{W}_\mathrm{RF})^{-1}\approx \frac{1}{N_\mathrm{r}}\mathbf{I}_{N_\mathrm{RF}}$ into (\ref{eqn:MSE1}), we have the following optimization problem
\begin{equation}\label{pro:opt3}
\begin{array}{cl}
\displaystyle{\minimize_{{\mathbf{V}_\mathrm{U}}}} & \sum\limits_{k = 0}^{N - 1}\mathrm{tr} (\frac{\gamma}{N_\mathrm{r}\sigma^2} \mathbf{V}^H_\mathrm{U}{\mathbf{C}}^H_{k}{\mathbf{C}_{k}}\mathbf{V}_\mathrm{U} + {\mathbf{I}}_{N_\mathrm{s}})^{-1} \\
\mathrm{subject \; to} & \mathbf{V}^H_\mathrm{U}\mathbf{V}_\mathrm{U}= \mathbf{I}_{N_\mathrm{s}},
\end{array}
\end{equation}
which is in a similar formulation of the analog beamforming problem in Section \ref{PrecodingEVD} and thus  $\mathbf{V}_\mathrm{U}$ can also be obtained using the EVD-based algorithm.

It is worth noting that when $N_\mathrm{RF}>N_\mathrm{s}$, the optimization of $\mathbf{V}_\mathrm{U}$, $\mathbf{V}_\mathrm{RF}$ and $\mathbf{W}_\mathrm{RF}$ is related to the factor $\gamma$, which is now not a constant as that in the case of $N_\mathrm{RF}= N_\mathrm{s}$. Thus, there should be some iterations between the optimization of $\mathbf{V}_\mathrm{D}$ (in turn $\gamma$) and that of $\mathbf{V}_\mathrm{RF}$ and $\mathbf{W}_\mathrm{RF}$, which complicates the overall algorithm. Nevertheless, in the simulation we found that an efficient way without the iteration is to directly set $\gamma = \frac{1}{N_\mathrm{t}N_\mathrm{s}}$ (the constant in the case of $N_\mathrm{RF}=N_\mathrm{s}$) in (\ref{JLB}) and (\ref{pro:opt3}) for the optimization of $\mathbf{V}_\mathrm{RF}$ and that of $\mathbf{V}_\mathrm{U}$, respectively, and finally set $\gamma$ as a power normalizing factor, i.e., $\gamma = \frac{1}{\mathrm{tr}{(\mathbf{V}_\mathrm{RF}\mathbf{V}_\mathrm{U}\mathbf{V}_\mathrm{U}^H \mathbf{V}_\mathrm{RF}^H})}$ to obtain the optimized $\mathbf{V}_\mathrm{D}=\sqrt{\gamma} \mathbf{V}_\mathrm{U}$. As a summary, the overall process of our proposed HBF scheme, which is referred to as EVD-HBF, is displayed in Algorithm 1.

\begin{figure}[t]
\begin{minipage}[b]{.45\linewidth}
  \centering
  \centerline{\includegraphics[width=4.2cm]{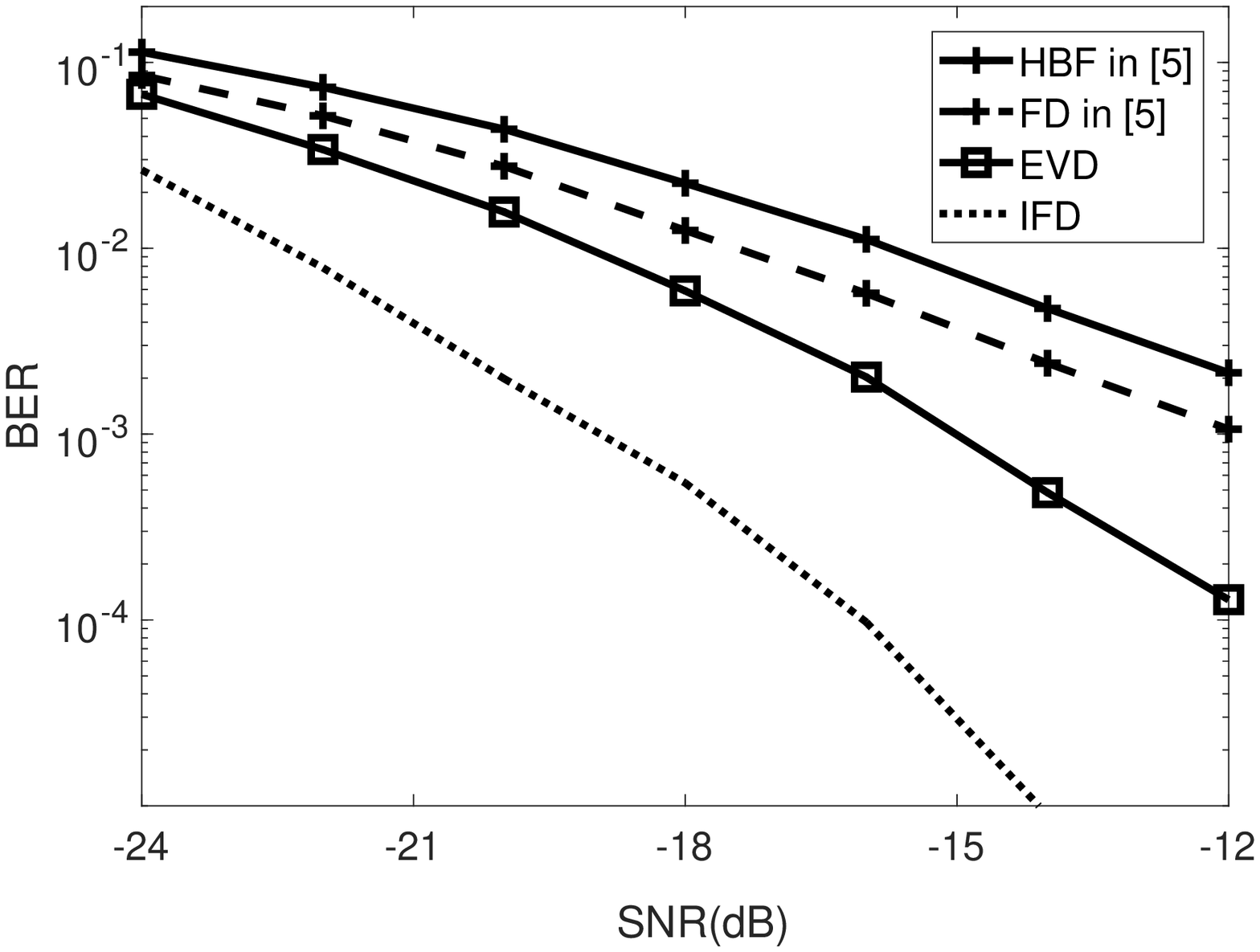}}
  \centerline{(a) }
\end{minipage}
\hfill
\begin{minipage}[b]{0.45\linewidth}
  \centering
  \centerline{\includegraphics[width=4.2cm]{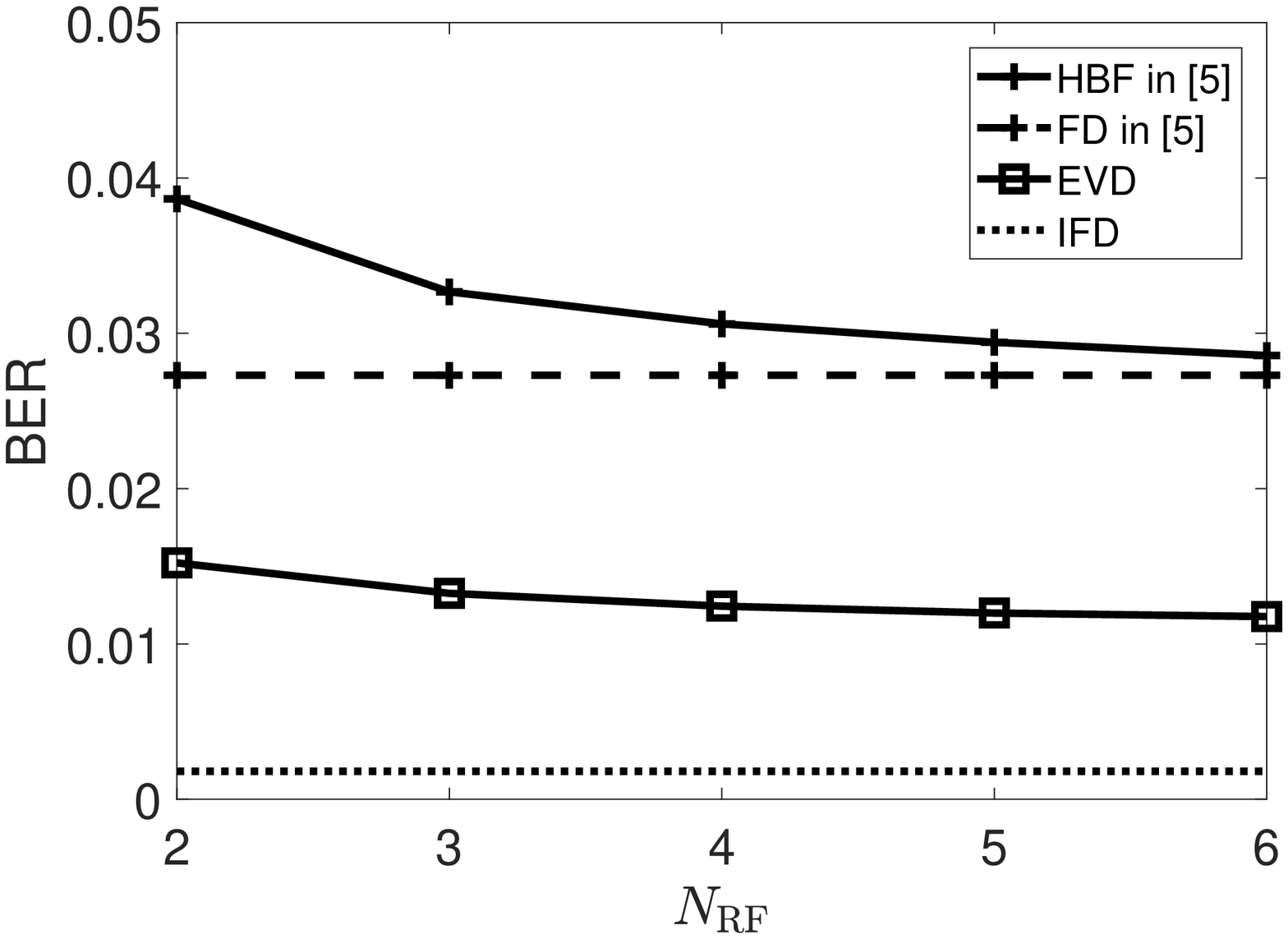}}
  \centerline{(b) }
\end{minipage}
\caption{Comparison of different beamforming schemes in a $64\times 64$ broadband SC mmWave MIMO system. (a) BER v.s. SNR. (b) BER v.s. $N_\mathrm{RF}$ when $\mathrm{SNR} = -18\mathrm{dB}$   and $N_\mathrm{s}$ = 2.}
\label{fig:res}
\end{figure}

\section{Simulation Results}
In the simulation, the mmWave channel samples were generated from  a geometry-based spatial channel model with $N_\mathrm{C}$ clusters and $N_\mathrm{R}$ rays within each cluster \cite{Li2017}, where the MIMO channel matrix  is represented by
\begin{equation}\label{eqn:Hmodel-broad}
\mathbf{H}(\tau) = \sqrt{N_\mathrm{t}N_\mathrm{r}}\sum_{i=1}^{N_{\mathrm{C}}} \sum_{m=1}^{N_{\mathrm{R}}} \alpha_{ij}{\mathbf{a}_\mathrm{r}}({\theta_{ij}^\mathrm{r}}){\mathbf{a}_\mathrm{t}}
({\theta_{ij}^\mathrm{t}})^H\cdot\delta(\tau-\tau_i), \nonumber
\end{equation}
where $\alpha_{ij}$ denotes the complex gain of the $j$th ray in the $i$th cluster, $\mathbf{a}_\mathrm{r}(\theta_{ij}^\mathrm{r})$ and $\mathbf{a}_\mathrm{t}{(\theta_{ij}^\mathrm{t})}$ denote the normalized responses of the transmit and receive antenna arrays to the $j$th ray in the $i$th cluster, respectively, with $\theta_{ij}^\mathrm{r}$ and $\theta_{ij}^\mathrm{t}$ denoting the angles of arrival and departure. $\delta(\cdot)$ is the dirac delta function, and $\tau_i$ denotes the delay of the $i$th cluster. More detail parameter settings can be referred to \cite{Li2017}.

Fig. 2 (a) shows the bit error ratio (BER) performance as a function of signal to noise ratio (SNR)  for different beamforming schemes with 4QAM when $N_\mathrm{t}=N_\mathrm{r}=64$ and $N_\mathrm{s}=N_\mathrm{RF}=2$. For comparison, we also provide the performance of two beamforming schemes in \cite{Buzzi2018}, i.e., the full-digital (FD) scheme and HBF scheme with FDE.  As a performance benchmark, we also provide the performance of an ideal FD scheme (labeled as 'IFD') based on the MMSE criterion \cite{S2001}. The difference between the FD scheme in \cite{Buzzi2018} and IFD in \cite{S2001} is that the former employs only one digital precoder for the whole band in order to guarantee a low PAPR while the latter employs an optimized precoder at each frequency tone without considering the PAPR effect. It can be seen from this figure that our proposed EVD-HBF scheme significantly outperforms both the conventional FD and HBF schemes for SC mmWave MIMO systems. This is because these schemes were optimized only based on the channel path with the largest gain.

Fig. 2 (b) shows the BER performance as a function of $N_\mathrm{RF}$ when $N_\mathrm{s}=2$ and $\mathrm{SNR}=-18\mathrm{dB}$. It can be seen that the proposed EVD-HBF scheme still outperforms the conventional FD and HBF schemes as $N_\mathrm{RF}$ increases and approaches to the IFD scheme.

\section{Conclusion}\label{conclusion}
We have studied the HBF design for broadband SC mmWave MIMO systems. Our basic idea was to first optimize the digital beamformers and then obtain a simplified problem containing only the analog beamforming variables. With some mathematical operations to reduce the difficulty, the analog precoder and combiner can be optimized alternatively via the EVD operation. Simulation results have demonstrated the significant improvement of the proposed EVD-HBF scheme.\\

\begin{center}
\textbf{{\normalsize  {ACKNOWLEDGMENT}}}
\end{center}
The authors would like to  thank the students Mr. Tianyi Lu and Ms. Anqi Jiang at Fudan University for their help in the proof of Theorem 1.

\bibliographystyle{IEEEbib}
\bibliography{SIP}

\end{document}